\documentclass[twocolumn,showpacs,aps,epsfig,nofootinbib]{revtex4}

\usepackage{graphicx}
\usepackage{epstopdf}
\usepackage{latexsym}
\usepackage{amssymb}
\usepackage{amssymb}

\usepackage[center]{subfigure}

\begin{document}

 \newcommand{\bq}{\begin{equation}}
 \newcommand{\eq}{\end{equation}}
 \newcommand{\bqn}{\begin{eqnarray}}
 \newcommand{\eqn}{\end{eqnarray}}
 \newcommand{\nb}{\nonumber}
 \newcommand{\lb}{\label}
\newcommand{\PRL}{Phys. Rev. Lett.}
\newcommand{\PL}{Phys. Lett.}
\newcommand{\PR}{Phys. Rev.}
\newcommand{\CQG}{Class. Quantum Grav.}

\title{U(1)  symmetry and elimination of spin-0 gravitons  in    Horava-Lifshitz   gravity without the projectability condition}

\author{Tao Zhu}  

\author{Qiang Wu}  

\author{Anzhong Wang  \footnote{On Leave from GCAP-CASPER, Physics Department, Baylor
University, Waco, TX 76798-7316, USA}} 

\affiliation{Institute  for Advanced Physics $\&$ Mathematics,   
Zhejiang University of
Technology, Hangzhou 310032,  China}

\author{Fu-Wen Shu}

\affiliation{College of Mathematics $\&$ Physics, Chongqing University of Posts  and Telecommunications, Chongqing 400065, China}

\date{\today}

\begin{abstract}

In this paper, we show that the spin-0 gravitons  appearing in Horava-Lifshitz  gravity without the projectability condition can be eliminated 
by extending the gauge symmetries of the foliation-preserving diffeomorphisms to include a local U(1) symmetry. As a result, the problems 
of stability, ghost, strong coupling,  and different speeds in the gravitational sector are automatically resolved. In addition, with the detailed 
balance condition softly breaking, the number of independent coupling constants can be significantly reduced (from more than 70 down to 
15), while  the theory is still UV complete and possesses  a healthy IR limit, whereby the prediction powers of the theory are considerably 
improved.  The strong coupling problem in the matter sector can be cured  by introducing an energy scale $M_{*}$, so that $M_{*} < 
\Lambda_{\omega}$, where $M_{*}$ denotes the suppression energy of  high order derivative terms, and $\Lambda_{\omega}$  the would-be 
strong coupling energy scale.

\end{abstract}


\maketitle

\section{ Introduction}

Quantization of  gravitational fields has been one of the main driving forces in Physics in the past  decades by following several 
different paths  \cite{QGs}.  Recently,  Horava proposed  a theory of quantum gravity in the framework of quantum field theory, with the spacetime metric as the elementary field
in the  language of the standard  path-integral  formulas \cite{Horava}. Applied to cosmology,  it results in various remarkable features and has attracted a 
great deal of attention \cite{reviews}.   In particular, high order spatial derivative terms can give rise to a bouncing universe \cite{Calcagni,brand};
the anisotropic scaling solves the horizon problem and leads to scale-invariant perturbations without inflation \cite{Muka,KKa}; the
lack of the local Hamiltonian constraint leads to ``dark matter as an integration constant" \cite{Mukb}; the dark sector can have its purely geometric origins \cite{Wang};
and the inclusion of a U(1) symmetry (with the projectability
condition) \cite{HMT,Silva} not only eliminates the spin-0 gravitons but also leads to a flat universe  \cite{HW}. 

With the perspective that Lorentz symmetry may appear only as an emergent one at low energies, but can be fundamentally
absent at high energies,  Horava considered a gravitational system whose scaling at short distances exhibits a strong anisotropy between space and time,
\bq
\lb{1.1}
{\bf x} \rightarrow b^{-1} {\bf x}, \;\;\;  t \rightarrow b^{-z} t.
\eq
This is quite similar to Lifsitz's  scalar field \cite{Lifshitz}, so the theory is often referred to as the Horava-Lifshitz (HL) gravity. 
In $(d+1)$-dimensions, in order for the theory to be power-counting renormalizable,   the critical exponent $z$  must  be $z \ge d$ \cite{Horava,Visser}.
The associated gauge symmetry   now is broken from the general diffeomorphisms down to  the  foliation-preserving
ones,  
\bq
\lb{1.4}
\delta{t} = - f(t),\; \;\; \delta{x}^{i}  =   - \zeta^{i}(t, {\bf x}),
\eq
to be denoted  by Diff($M, \; {\cal{F}}$). 

Abandoning  the general diffeomorphisms, on the other hand,   gives rise  to   a proliferation of independently  coupling constants
\cite{Horava,BPS}, which  could potentially limit   the prediction powers of the theory. In particular, only the sixth-order spatial derivative terms are more than 60 \cite{KP}. %
To reduce the number of the independent coupling constants,  two conditions were introduced, the {\em projectibility} and {\em detailed balance} \cite{Horava}. 
The former requires that the lapse function $N$ be a function of $t$ only, while the latter requires that
the gravitational potential  should be obtained from a superpotential $W_{g}$,
where  $W_{g}$  is given by an integral of  
the gravitational Chern-Simons term over a 3-dimensional space, $W_{g}  \sim \int_{\Sigma}{\omega_{3}(\Gamma)}$. 
With these two conditions, the general action now contains only five  independent  coupling constants   \cite{Horava}.  The detailed balance condition has   
several   remarkable features \cite{Horava,BLW}.  For example,  it is in the same spirit of the AdS/CFT correspondence \cite{AdSCFT}, where a string theory and 
gravity defined on one space is  equivalent to a quantum field theory without gravity defined on the conformal boundary of this space,  which has one or more  lower  
dimension(s).  Yet,  in the non-equilibrium  thermodynamics,  the counterpart of the superpotential $W_{g}$  plays the role of entropy, while   ${\delta{W}_{g}}/{\delta{g}_{ij}}$ 
represents  the corresponding entropic force \cite{OM}. This might shed  further lights on the nature of gravitational forces, as proposed recently by Verlinde \cite{Verlinde}.  

Despite of all the above remarkable features, the theory (with these two conditions) 
is plagued with several problems \cite{reviews}, including the instability, ghost 
\cite{Horava,Ins}, and strong coupling \cite{SC,KP}.  However, all of them are closely related to the existence of the spin-0 gravitons \cite{reviews}. Because of their
presence,  another  important  question   also raises:  Their speeds are generically different from those of gravitational waves, as  they are not related by any symmetry. This poses 
a great challenge for any attempt to restore Lorentz symmetry  at low energies where it has been  well tested experimentally.  In particular, one needs a mechanism 
to ensure that in those energy scales all species, including 
gravitons,  have the same effective speed and light cones.  With these in mind,   
 Horava and Melby-Thompson (HMT)     \cite{HMT}  extended the  symmetry  (\ref{1.4})     to include  a local $U(1)$,
\bq
\lb{symmetry}
 U(1) \ltimes {\mbox{Diff}}(M, \; {\cal{F}}).
 \eq
With this enlarged symmetry, the spin-0 gravitons are eliminated \cite{HMT,WW,Kluson},  and the  theory has the same long-distance limit as that of general relativity.
This was initially done in the special case  $\lambda =1$,   and soon generalized to the case
with any $\lambda$  \cite{Silva}, where $\lambda$ is a coupling constant defined below. Although the spin-0 gravitons are also eliminated in the general case, 
 the strong coupling problem raises again
 \cite{HW}. However, it can be solved by the  Blas-Pujolas-Sibiryakov (BPS) mechanism \cite{BPS}, in which   a new energy scale $M_{*}$ is introduced, 
 so that $M_{*} < \Lambda_{\omega}$,  where $M_{*}$ denotes the suppression energy of  high order derivative terms of the theory, and 
$\Lambda_{\omega}$  the would-be strong coupling energy  scale \cite{LWWZ}. 

Note that in \cite{HMT,Silva} 
 the projectability condition was assumed.  In this paper, we shall study the case
 without it, and our goals are twofold: (i) Extend the symmetry (\ref{symmetry}) to the case without the projectability condition, so that the spin-0 gravitons are eliminated.
(ii) Reduce significantly the number of the coupling constants by the detailed balance condition, whereby the prediction powers of the theory can be improved
considerably.  To have a healthy IR, we allow it to be  broken softly by adding all the  low dimensional relevant terms 
\cite{BLW}. We also note that in  \cite{Klusonb}  a  U(1) extension of F(R) HL gravity was  recently studied.

 \section{The Model}

Under the U(1) transformations, the metric coefficients transform as \cite{HMT},  
\bq
\lb{gauge}
 \delta_\alpha N = 0, \;\;\; \delta_\alpha N_i=N\nabla_i\alpha,\;\;\; \delta_\alpha g_{ij}=0,
\eq
where $\alpha$ is the U(1) generator, $N_{i}$ the shift vector, and $\nabla_i$ the covariant derivative 
of $g_{ij}$. Since \cite{BPS}
\bq
\lb{GR}
\hat{S}_{g} = \zeta^{2} \int{dt dx^{3}N\sqrt{g}\left({\cal{L}}_{K}  - {\cal{L}}_{V}(g_{ij}, a_{k})\right)}, 
\eq
where  $a_{i} \equiv (\ln N),_{i}, $
${\cal{L}}_{K} \equiv 
K_{ij}K^{ij} - \lambda K^{2}$ and  $K_{ij} =\left(- \dot{g}_{ij} + \nabla_{i}N_{j} + \nabla_{j}N_{i}\right)/(2N)$,  
  the potential ${\cal{L}}_{V}$ is invariant under (\ref{gauge}),    while the kinetic part
transforms as
\bqn
\lb{actiontrans}
&& \delta S_K =  \zeta^2\int dtd^3x N \sqrt{g}\Big\{\big(\dot{\alpha}-N^i\nabla_i\alpha\big)\frac{R}{N} + 2\alpha G_{ij} K^{ij} \nb\\
& &  + 2K_{ij}\hat{\cal G}^{ijlk}a_{(l}\nabla_{k)}\alpha 
+  2(1-\lambda) K\big(\nabla^2\alpha+a^k\nabla_k\alpha\big)\Big\}, ~~~
\eqn
where $\hat{\cal G}^{ijkl} = \left.{\cal G}^{ijkl}\right|_{\lambda =1}$,  ${\cal G}^{ijkl}(\lambda) \equiv (g^{ik}g^{jk}+ g^{il}g^{jk})/2 -\lambda g^{ij}g^{lk}$, 
$G_{ij} \equiv R_{ij} - Rg_{ij}/2$ and $f_{(ij)} = (f_{ij} + f_{ji})/2$. $R_{ij} \; (R)$ is the Ricci tensor (scalar)  of $g_{ij}$. 
To have the $U(1)$ symmetry,   we introduce the gauge field $A$  \cite{HMT}, which transforms as
\bq
\lb{eqa}
\delta_\alpha A=\dot{\alpha}-N^i\nabla_i\alpha.
\eq
Then, adding \footnote{Note the difference between the notations used here and the ones used in \cite{HMT,Silva}. In particular, we
have $K_{ij} = - K_{ij}^{HMT},\; \Lambda_{g} = \Omega^{HMT},\; \varphi = - \nu^{HMT}, {\cal{G}}_{ij} = \Theta_{ij}^{HMT}$, 
where quantities with the superindices ``HMT" were  used in \cite{HMT,Silva}.} 
\bq
\lb{actionA}
S_A=-\zeta^2\int dtd^3x\sqrt{g}A(R-2\Lambda_{g}),
\eq
to $\hat{S}_{g}$, one finds that its variation (with $\Lambda_{g} = 0$) with respect to $\alpha$ exactly cancels the first term in Eq.(\ref{actiontrans}).  
To repair the rest,   we introduce  the Newtonian prepotential $\varphi$  \cite{HMT}, which transforms as 
\bq
\lb{eqb}
\delta_\alpha\varphi= - \alpha.
\eq
Under Eq.(\ref{gauge}),  the variation of  the term,   
\bqn
\lb{action nu}
&& S_\varphi=\zeta^2\int dtd^3x\sqrt{g}N\Big\{\varphi{\cal{G}}^{ij}\big(2K_{ij}+\nabla_i\nabla_j\varphi+a_i\nabla_j\varphi\big)\nb\\
& &~ +(1-\lambda)\Big[\big(\nabla^2\varphi+a_i\nabla^i\varphi\big)^2 
+2\big(\nabla^2\varphi+a_i\nabla^i\varphi\big)K\Big]\nb\\
& &~ +\frac{1}{3}\hat{\cal G}^{ijlk}\Big[4\left(\nabla_{i}\nabla_{j}\varphi\right) a_{(k}\nabla_{l)}\varphi + 5 \left(a_{(i}\nabla_{j)}\varphi\right) a_{(k}\nabla_{l)}\varphi\nb\\
&& ~
+ 2 \left(\nabla_{(i}\varphi\right)a_{j)(k}\nabla_{l)}\varphi
+ 6K_{ij} a_{(l}\nabla_{k)}\varphi
\Big]\Big\},
\eqn
will exactly cancel the rest in Eq.(\ref{actiontrans}) as well as the one from the term $\Lambda_{g} A$ in Eq.(\ref{actionA}),
where ${\cal{G}}_{ij} \equiv G_{ij}  + \Lambda_{g} g^{ij}$ and $a_{ij} \equiv \nabla_{i}a_{j}$. Hence, the total action
\bq
\lb{actions}
S_{g} = \hat{S}_{g} + S_{A} + S_{\varphi},
\eq
 is invariant under (\ref{symmetry}). 
%
  
Equation (\ref{symmetry}) does not uniquely fix ${\cal{L}}_{V}$. To our second  goal, we
 introduce  the ``generalized" detailed balance condition,
\bqn
\lb{2.4}
\hat{\cal{L}}_{V} &=& E_{ij}{\cal{G}}^{ijkl}E_{kl}  - g_{ij}A^{i}A^{j},
\eqn
where 
\bqn
\lb{2.5a}
E^{ij} &=& \frac{1}{\sqrt{g}}\frac{\delta{W}_{g}}{\delta{g}_{ij}},\;\;\;
A^{i} =  \frac{1}{\sqrt{g}}\frac{\delta{W_{a}}}{\delta{a_{i}}},\nb\\
W_{g}  &=& \frac{1}{w^{2}}\int_{\Sigma}{ {\mbox{Tr}} \Big(\Gamma \wedge d\Gamma + \frac{2}{3}\Gamma\wedge \Gamma\wedge \Gamma\Big)},\nb\\
W_{a} &=& \int{d^{3}x \sqrt{g} a^{i}\sum_{n= 0}^{1}{b_{n}\Delta^{n}{a_{i}}}},
\eqn
with   $\Delta\equiv g^{ij}\nabla_{i}\nabla_{j}$ and $b_{n}$ being arbitrary  constants. Now, some comments are in order. 
First, 
 the term  $a_{i}\Delta^{1/2}a^{i}$ in principle can  be included into $W_{a}$, which will give rise 
to fifth order derivative terms. In this paper we shall discard   them by   parity. Second, it is well-known that with the detailed 
balance condition a scalar field is not UV complete \cite{Calcagni}, and
the Newtonian limit does not exist \cite{LMP}. Following \cite{BLW},  we break it softly by adding all the low dimensional relevant terms to 
$\hat{\cal{L}}_{V}$, so that the potential finally takes the form, 
\bqn
\lb{potential}
{\cal{L}}_{V} &=&  \gamma_{0}\zeta^{2}  -  \Big(\beta_{0}  a_{i}a^{i}- \gamma_{1}R\Big)
+ {\zeta^{-2}} \Big(\gamma_{2}R^{2} +  \gamma_{3}  R_{ij}R^{ij}\Big)\nb\\
& & + {\zeta^{-2}}\Big[\beta_{1} \left(a_{i}a^{i}\right)^{2} + \beta_{2} \left(a^{i}_{\;\;i}\right)^{2}
+ \beta_{3} \left(a_{i}a^{i}\right)a^{j}_{\;\;j} \nb\\
& & + \beta_{4} a^{ij}a_{ij} + \beta_{5}
\left(a_{i}a^{i}\right)R + \beta_{6} a_{i}a_{j}R^{ij} + \beta_{7} Ra^{i}_{\;\;i}\Big]\nb\\
& &   
 +  {\zeta^{-4}}\Big[\gamma_{5}C_{ij}C^{ij}  + \beta_{8} \left(\Delta{a^{i}}\right)^{2}\Big],
\eqn
where $ \beta_{0} \equiv b_{0}^{2}, \; \gamma_{5} \equiv \zeta^{4}/w^{4},\; \beta_{8} \equiv - \zeta^{4}b_{1}^{2}$, and
$C_{ij}$ denotes the Cotton tensor, defined as
$
C^{ij} \equiv \frac{{{e}}^{ikl}}{\sqrt{g}}\nabla_{k}\Big(R^{j}_{l} - \frac{1}{4}R\delta^{j}_{l}\Big)$. 
 All the coefficients, $ \beta_{n}$ and $\gamma_{n}$, are
dimensionless and arbitrary, except $\beta_{0} \ge 0, \gamma_{5} \ge 0$ and $\beta_{8}\le 0$, as indicated by their definitions.
 $\gamma_{0}$ is related to the cosmological constant by $\Lambda =  \zeta^{2}\gamma_{0}/2$, while
the IR limit requires 
$\gamma_{1} = -1, \; \zeta^{2} = 1/(16\pi G)$, where $G$ is the Newtonian constant. 
 From the above, we can see that with the ``generalized" detailed balance  condition softly breaking, the number of independent coupling constants is significantly reduced
 from more 70  to 15:  $G, \; \Lambda, \; \lambda, \; \beta_{n},\; \gamma_{s},\; (n = 0, ..., 8; \; s = 2, 3, 5)$.
  In the following, we shall show that such a setup 
  is UV complete and IR healthy. 

\section{Elimination of the Spin-0 Gravitons}


To show this, it is sufficient to consider linear perturbations in the Minkowski background, given by
\bqn
\lb{pertbs}
N&=& 1+\phi, \;\;\; N_i=\partial_i B,\;\;\;  g_{ij}=(1-2\psi)\delta_{ij}+ 2E_{,ij},\nb\\
 A&=& \delta A, \;\;\; \varphi =\delta \varphi.
\eqn
Choosing the gauge, $E=0 = \delta\varphi$,  we find that
\bqn
\lb{second action}
S_{g}^{(2)}&=&\zeta^2\int dtd^3x\Big\{(1-3\lambda)(3\dot{\psi}^2+2\dot{\psi}\partial^2B)\nb\\
& &+(1-\lambda)(\partial^2B)^2 -\left(\phi\eth+ {4\beta_7}{\zeta^{-2}}\partial^2\psi\right)\partial^2\phi\nb\\
&&-2(\psi-2\phi+2A+\alpha_1\psi\partial^2)\partial^2\psi\Big\},
\eqn
where $\alpha_{1} \equiv \zeta^{-2}(8\gamma_{2} + 3\gamma_{3})$ and $\eth  \equiv \beta_{0} + \zeta^{-2}(\beta_{2}+\beta_{4})\partial^{2}   
 -\zeta^{-4} \beta_{8}\partial^4$.  Variations of $S^{(2)}_{g}$ with respect to $A$, $\psi$, $B$, and $\phi$ yield, respectively,
\bqn
\lb{A}
&&\partial^2\psi=0,\\
\lb{psi}
&& \ddot{\psi} + \frac{1}{3}\partial^{2}\dot{B} + \frac{2\partial^{2}(A+\psi + \alpha_{1}\partial^{2}\psi)}{3(1-3\lambda)} 
 =  \frac{2\wp\partial^{2}\phi}{3(1-3\lambda)},~~~~~~\\
\lb{B}
&&\big(\lambda - 1\big)\partial^{2}B = \big(1-3\lambda\big)\dot{\psi},\\
\lb{phi}
&&\eth\phi=2\wp\psi,
\eqn
where  $\wp \equiv 1  -  {\zeta^{-2}} {\beta_{7}}\partial^{2}$. Eq.(\ref{A}) shows that $\psi$ is not propagating, and with proper boundary conditions, one can  set $\psi=0$. Then, 
Eqs.(\ref{psi})-(\ref{phi}) show that $B$, $A$, and $\phi$  are also not propagating and can be  set to zero. 
Hence, we obtain $\psi=B=A=\phi=0$, that is, the scalar perturbations vanish identically in the Minkowski background, similar to that in general relativity. Thus, with the enlarged symmetry
(\ref{symmetry}),  the spin-0 gravitons are indeed eliminated even without the projectability condition.

\section{Strong Coupling and the  Blas-Pujolas-Sibiryakov Mechanism}

Since the spin-0 gravitons are eliminated, the ghost, instability,  strong coupling and different speed problems in the gravitational sector 
do not exist any longer. But, the self-interaction of matter fields and the interaction between a matter field and a gravitational field can still lead to strong coupling, as shown recently in \cite{LWWZ} for the
 theory with the projectability condition. In the following we shall show that   this is also the case here. Since the proof is quite similar to that given in \cite{LWWZ}, in the following we just summarize 
our main results, and for detail, we refer readers to \cite{LWWZ}. 
A scalar field $\chi$ with detailed balance condition softly breaking is described by    Eqs.(3.11) and (3.12) of \cite{BLW}. In the Minkowski background, 
we have $\bar\chi = 0 = V(0) =V'(0)$. Considering the linear perturbations $\chi = \delta\chi$, we find that the quadratic part of the scalar field action reads,
\bqn
\lb{eqq1}
&& 
S_{\chi}^{(2)} = \int dtd^3x\Bigg[\frac{1}{2}{f}\dot{\chi}^2- \frac{1}{2}V''\chi^2 -\frac{1}{2}\left(1+2V_1\right)(\partial\chi)^2\nb\\
&&
+c_1A\partial^2\chi -V_2(\partial^2\chi^2)^2 
-V_4'\chi\partial^4\chi+\sigma_3^2\partial^2\chi\partial^4\chi\Bigg],
\eqn
where $f$ is a constant, usually chosen to be one \cite{BLW}. 
Since  $S_{\chi}^{(2)} $ does not depend on $\psi,\; B$ and $\phi$ explicitly, the variations of the total action $S^{(2)} = S_{\chi}^{(2)} + S_{g}^{(2)}$ with respect to them
will give the same Eqs.(\ref{psi}) - (\ref{phi}),
while Eq.(\ref{A}) is replaced by $\psi = c_1\chi/(4\zeta^{2})$. Integrating out $\phi,\; \psi,\; B, \; A$, we obtain
\bqn
\lb{3eqq2}
&& S^{(2)} = M^{2}\int{dtd^{3}x\Bigg[\dot\chi^{2} - \alpha_0\big(\partial\chi\big)^{2} - m_{\chi}^{2} \chi^{2} }\nb\\
&& - {M_{A}^{-2}} \chi\partial^{4}\chi
        + {M_{B}^{-4}} \chi\partial^{6}\chi
            +\gamma\chi\partial^2\left(\frac{\wp^2\chi}{\eth}\right)\Bigg],
\eqn
where $M^{2} \equiv {2\pi G c_{1}^{2}}/{|c_{\psi}|^{2}}   + {f}/{2}$, $\gamma  \equiv {4\pi Gc_1^2}/M^2$, 
$\alpha_0 \equiv\big(1 + 2V_{1} - 4\pi G c_{1}^{2}\big)/(2M^{2})$,  
$M^{2}_{A} \equiv M^{2}/\big(2\pi G \alpha_{1}c_{1}^{2}   + V_{2} + V_{4}'\big)$, 
$M^{4}_{B} \equiv M^{2}/\sigma_{3}^{2},\; m^{2}_{\chi} \equiv V''/(2M^{2})$, 
and $c_{\psi}^{2} = (1-\lambda)/(3\lambda - 1)$. Clearly, the scalar field is ghost-free for $f >0$, and stable in the UV and IR. In fact, it can be made  stable in all energy scales 
by properly choosing the coupling coefficients $V_{n}$.  

To consider the strong coupling problem, one needs to calculate the cubic part of the total action, which takes the form,
 \bqn
 \lb{3.10a}
&& S^{(3)} =  \int{dtd^{3}x\Bigg\{\lambda_{1}\left(\frac{1}{\partial^{2}}\ddot{\chi}\right)\chi\partial^{2}\chi  + \lambda_{2}\left(\frac{1}{\partial^{2}}\ddot{\chi}\right)\chi_{,i}\chi^{,i}}
 \nb\\
  && +  \lambda_{3} \dot{\chi}^{2}\left(\frac{2\wp }{\eth} -1\right)\chi   +  \lambda_4 \dot{\chi}\partial^i\left(\frac{2\wp }{\eth} -1\right)\chi\partial_{i}\left(\frac{\dot{\chi}}{\partial^{2}}\right)\nb\\
   &&
 + \lambda_5 \left(\frac{\partial^{i}\partial^{j}}{\partial^{2}}\dot\chi\right)\left(\frac{\partial_{i}}{\partial^{2}}\dot{\chi}\right)\partial_j\left(\frac{2\wp }{\eth} +3\right)\chi
  \nb\\
  &&
+  \lambda_6 \dot\chi \chi^{,i}\left(\frac{\partial_{i}}{\partial^{2}}\dot\chi\right)
  + ...\Bigg\}, 
 \eqn
 where ``..." represents the terms that   are independent of $\lambda$, so they are irrelevant to the strong coupling problem.
 $\lambda_{s}$ are functions of $\lambda, f, \zeta$ and $c_{i}$. In particular, $\lambda_{5} =  {c_1^3}/{(64\zeta^4 |c_\psi|^{4})}$, which will
yield the strong coupling energy $\Lambda_{\omega}$ given below. The exact expressions of other coefficients
 are not relevant to $\Lambda_{\omega}$, so  will not be given here. For a process with energy $E \ll M_{A}, M_{B}$, the first two terms in Eq.(\ref{3eqq2}) 
 are dominant, and $ S^{(2)}$ is invariant under the relativistic rescaling
$t \rightarrow b^{-1} t, x^{i} \rightarrow b^{-1} x^{i}, \chi \rightarrow b\chi$. Then, all the terms of $\lambda_{s}$ in $S^{(3)}$ scale as $b$.
As a result, when the energy of a process is greater than a certain value, say, $\Lambda_{\omega}$,
 the amplitudes of these terms are greater than one, and the theory becomes nonrenormalizable \cite{Pol}. In the present case, it can be shown that
 \bq
 \lb{SC}
 \Lambda_{\omega} \simeq \left(\frac{M_{pl}}{c_1}\right)^{3/2}M_{pl}|c_{\psi}|^{5/2}. 
 \eq
 However, if 
 \bq
 \lb{SCa}
 M_{*} <  \Lambda_{\omega},
 \eq
 where $M_{*} = {\mbox{Min.}}(M_{A}, M_{B})$, one can see that before the strong coupling energy  $\Lambda_{\omega}$ is reached, the high order derivative
 terms in Eq.(\ref{3eqq2}) become large, and their effects must be taken into  account. In particular, for $M_{A} \gtrsim M_{B}$, the sixth-order derivative
 terms become dominant for  processes with   $E \gtrsim M_{*}$,    and the quadratic action (\ref{3eqq2}) now is invariant  only under the anisotropic  rescaling, 
 \bq
 \lb{ASC}
 t \rightarrow b^{-3} t, \;\;\; x^{i} \rightarrow b^{-1} x^{i},\;\;\; \chi \rightarrow \chi.
 \eq
 Then, one finds that under the new rescaling all the terms of $\lambda_{s}$ in Eq.(\ref{3.10a}) become scaling-invariant, while the rest  scale as $b^{-\delta}$ with $\delta > 0$. The former 
 are strictly renormalizable, while the latter 
 are superrenormalizable \cite{Pol}. Therefore  Eq.(\ref{SCa}) makes the strong coupling problem disappeared.
 
\section{Conclusions}

 In this paper, we have shown that the spin-0 gravitons in the HL theory even without the projectability condition can be eliminated by the enlarged symmetry 
 (\ref{symmetry}). 
 An immediate result is that all the problems related to them  disappear, including
 the ghost, instability, strong coupling and different speeds in the gravitational sector. In addition, the requirements \cite{BPS}, $|\lambda -1| \simeq \beta_{0},\; 0 < \beta_{0} < 2$, now become unnecessary, which might help 
 to  relax the observational constraints. Moreover, it is  exactly because of this elimination that softly breaking detailed balance condition can be imposed, whereby
   the number of independent coupling constants is significantly reduced  from more than 70 to 15. This   considerably improves the prediction powers of the theory. 
 Note that,  without the elimination of the spin-0 gravitons, the strong coupling problem will appear in the gravitational sector and  cannot be  solved by the BPS mechanism, because  this 
 condition prevents the existence of sixth-order derivative terms in $S^{(3)}_{g}$.  
 
 These results  put the HL theory with/without the projectability condition in the same footing, and provide a very promising 
 direction to build a viable theory of quantum gravity,   put forwards recently  by HMT \cite{HMT}. Certainly,  many challenging questions  \cite{reviews} 
 need  to be answered   before such a goal is finally reached. 
 

{\bf Acknowledgements:}   
This work was supported in part by DOE  Grant, DE-FG02-10ER41692 (AW);  NNSFC No. 11005165   (FWS);
and NNSFC No. 11047008 (QW, TZ).


\end{document}